# A "User Experience 3.0 (UX3.0)" Paradigm Framework: User Experience Design for Human-Centered AI Systems

Wei Xu

(Center for Psychological Sciences, Zhejiang University, Hangzhou 310058, China)

**Abstract**: The human-centered artificial intelligence (HCAI) design approach, the user-centered design (UCD) version in the intelligence era, has been promoted to address potential negative issues caused by AI technology; user experience design (UXD) is specifically called out to facilitate the design and development of human-centered AI systems. Over the last three decades, user experience (UX) practice can be divided into three stages in terms of technology platform, user needs, design philosophy, ecosystem, scope, focus, and methodology of UX practice. UX practice is moving towards the intelligence era. Still, the existing UX paradigm mainly aims at non-intelligent systems and lacks a systematic approach to address UX for designing and developing human-centered AI products and systems. The intelligence era has put forward new demands on the UX paradigm. This paper proposes a "UX 3.0" paradigm framework and the corresponding UX methodology for UX practice in the intelligence era. The "UX 3.0" paradigm framework includes four categories of emerging experiences in the intelligence era: ecosystem-based experience, innovation-enabled experience, AI-enabled experience, and human-AI interaction-based experience, each compelling us to enhance current UX practice in terms of design philosophy, scope, focus, and methodology. We believe that the "UX 3.0" paradigm helps enhance existing UX practice and provides methodological support for the research and applications of UX in developing human-centered AI systems. Finally, this paper looks forward to future work implementing the "UX 3.0" paradigm.

## 1 Introduction

The concept of user experience (UX) and "user-centered design (UCD)," proposed by Norman (1986), have driven UX practice and human-computer interaction ( HCI) in the computer era (Nielson, 1993; Xu, 2003, 2005). Over the past three decades, the UX field has grown from a narrow usability design to a holistic UX design approach, including design for total user experience, service experience, experience thinking, and experience management (Dong et al., 2021; Xin, 2019; Huang, Lai, 2020; Huang et al., 2022; Xu, 2017, 2018).

UX has entered a new era: the intelligent era and the popularization of AI-based intelligent products have brought new requirements and opportunities to a new stage of UX practice (Xu, 2019b, 2020a; Xu, Gao, Ge, 2022). For example, the design approach of human-centered artificial intelligence (AI) (HCAI), the UCD version in the intelligence era, has been promoted to address potential negative issues caused by AI technology (Xu, 2019; Shneiderman, 2020); user experience design (UXD) is specifically called out to develop HCAI systems. On the other hand, new characteristics of AI technology pose challenges to UXD, such as emerging user needs, autonomous features, intelligent user interface, human-AI interaction, and human-AI collaboration. Xu, Gao, and Dainoff (2024) proposed a methodological framework for implementing the HCAI approach, including design principle, implementation approach, process, method, and interdisciplinary team. However, no one has proposed a systematic UX paradigm to enhance UXD to support HCAI products in the intelligence era. Such challenges require us to reassess the status of UX practice: *What kind of UX paradigm, which can effectively guide UX practice, should be developed to enable UXD to develop HCAI products in the intelligence era?*

To answer this question, this paper first analyzes the cross-era evolution of UX practice and identifies the emerging requirements faced by UX practice in the intelligent era. Then, it proposes a "UX 3.0" paradigm framework for UX practice in the intelligence era based on our previous work. Finally, the paper looks forward to future work with the " UX 3.0 " paradigm. The purpose is to provide methodological support for more effective UX research and application in the intelligence era, enabling UXD to develop HCAI systems.

## 2 New requirements for the UX paradigm in the intelligence era
### 2.1 Cross-era evolution of UX practice

Over the last three decades, UX practice can be divided into three stages in terms of technology platform, user needs, design philosophy, ecosystem, scope, focus, and methodology of UX practice (see Table 1) (Xu, 2018). The "PC/Internet Era" ushered in the first "exploring" stages of UX practice. In 2007, Apple launched the iPhone, which innovated the mobile experience with mobile and Internet-based technologies. It marked the beginning of the "mobile Internet era" and initiated the second "growing" stage of UX practice. In 2015, Google's AlphaGo was launched, and AI-related technologies such as deep machine learning and big data have rapidly grown since then, marking the advent of the intelligence era. AI-based intelligent products enter people's daily work and lives, requiring more systematic and mature practice and marking the beginning of the third "maturing" stage of UX practice.

Table 1 Comparison of cross-era characteristics of UX practice

| Characteristics of UX practice | UX 1.0 (Exploring Stage) (Late1980s – ~ 2007 ) | UX 2.0 (Gowing Stage) ( ~ 2007- ~ 2015 ) | UX 3.0 (Maturing Stage) ( ~ 2015– ) |
|---|---|---|---|
| **Technological platform** | PC / Internet Era | Mobile Internet era | Intelligence era (big data, AI, metaverse…) |
| **User needs** | Product functionality, usability, etc. | + Total UX (just started) (Dong et al., 2016), information security, etc. | Total UX + intelligent, natural, personalized, and emotional HCI, human-AI collaboration, ethics and morality, privacy, decision-making authority, skill growth, etc. |
| **Design philosophy** | User-centered | User-centered | User-centered (including "human-centered AI ") (Xu, 2019b) |
| **UX ecosystem** | None | Start to form | Initially formed |
| **Focus** | Usability of the user interfaces of individual products (siloed solutions) | UX is no longer limited to usability but also includes business processes and other user interaction touchpoints (Xu, 2012) | End-to-end UX: ecosystem-based experience, innovation-enabled experience, AI-enabled experience, human-AI interaction-based experience. See detailed discussions below |
| **Scope** | Product development stage | Product life cycle (pre-development, development, post-development) | Product life cycle + macro intelligent socio-technical systems environment |
| **Methodology** | Usability engineering (Nielson, 1993): user research, UI prototyping, usability testing, etc. | UX-based approach (e.g., Xu, 2017): Beyond usability and user interface | End-to-end UX: See detailed discussions in Section 3 |

As shown in Table 1, the three UX stages across technological eras show prominent characteristics. These new characteristics also demonstrate the distinct UX paradigms representing the UX practice across the three stages. The paradigm of a field is the lens that frames the perspective of research and determines the scope, focus, and methodology of research and application for the field (Xu, Gao, Ge, 2024). As highlighted in Table 1, there are prominent characteristics throughout the history of UX practice. With the help of emerging technologies and user needs, UX paradigms have been developing and advancing UX practice.

Figure 1 further illustrates the evolution of the UX paradigm over the past 30 years. As shown in Figure 1, from a UX methodological perspective as an example, in the first UX stage, the UX paradigm was mainly based on usability engineering methods for user interface design and usability testing (Neilson, 1993). In the second stage, the UX paradigm begins to transcend the usability of the human-computer interface, and UX methods were developed or enhanced, for example, the concept of total UX (Finstad, Xu et al., 2009; Xu, 2014; Xu, 2017).

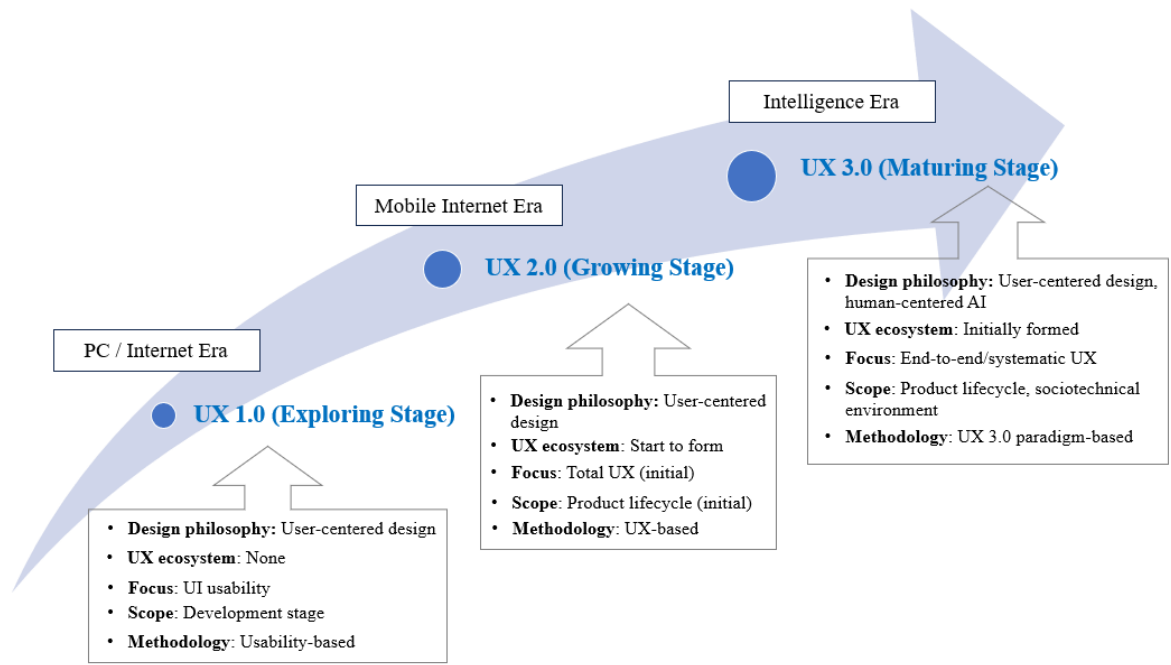

Figure 1 Evolution of UX paradigms across technological eras

As we enter the third stage (i.e. the intelligence era), UX practice faces new challenges. AI-based intelligent products benefit humans, but research shows that inappropriate design, development, and use of AI technology may affect UX and even harm humans. For example, the AI accident database has collected more than 1,000 accidents. These accidents include self-driving cars hitting and killing pedestrians, trading algorithm errors causing market "flash crashes," etc. These accidents will inevitably lead to a negative user experience (McGregor, 2023). Many AI developers primarily follow the "technology-centered" approach in development and believe that problems that could not be solved by human-computer interaction in the past have been solved by intelligent technology (such as voice input), so UX does not need to be solved anymore. Besides, UX professionals often participated in AI projects only after the product requirements were defined, limiting their influence on the design of intelligent systems and leading to the failure of some AI projects (Yang et al., 2020; Budiu & Laubheimer, 2018).

In response to these challenges, people have proposed the concept of "human-centered AI (HCAI)" (Li, 2018; Xu, 2019; Shneiderman, 2020). As one of the primary design goals, HCAI aims to optimize intelligent systems' interaction and experience design. UX plays an essential role in the success or failure of intelligent systems, such as the experience design of intelligent human-computer interfaces and iterative user interface prototyping/usability testing. Thus, the UX practice in the intelligence era is needed to guide UX practice, including design philosophy, focus, scope, and methodology to of UX practice, enabling and supporting rt the design and development of human-centered AI systems.

## 2.2 The conceptual framework for the "UX 3.0" paradigm

Researchers have begun to put forward some preliminary considerations of UX paradigms for UX practice in the intelligence era from a methodological perspective. For example, researchers use technologies such as AI and big data to model real-time online user behavior and other data to provide users with personalized designs (Tan et al., 2020; Herath & Jayarathne, 2018; Quadrana et al., 2017; Lu, Zhu, 2018). Xin (2019) proposed a new paradigm for transitioning from user experience to experience design. In the field of human factors science, researchers have begun to pay attention to UX issues related to intelligent systems (e.g., Stephanidis, Salvendy, et al., 2019; Dong et al., 2021; Shneiderman et al., 2016; Preece et al., 2019; Xu, Ge, 2020). To enable the design and development of HCAI products, Wei & Gao (2024) proposed a methodological framework for implementing the HCAI approach. However, no one has proposed a systematic UX paradigm to enhance UXD to support HCAI products in the intelligence era.

As discussed earlier, existing UX paradigms are mainly for UX practice for non-intelligent systems and lack a systematic approach for intelligent systems. On the one hand, AI technology presents new challenges to UX professionals who have entered the deep-water area of UX practice in the intelligence era; on the other hand, this also provides new opportunities for us to advance UX paradigms.

To meet the new requirements for UX paradigms in the intelligence era, based on our previous research and other's work, this paper proposes a conceptual framework for the "UX 3.0" paradigm in the intelligence era (see Figure 2).

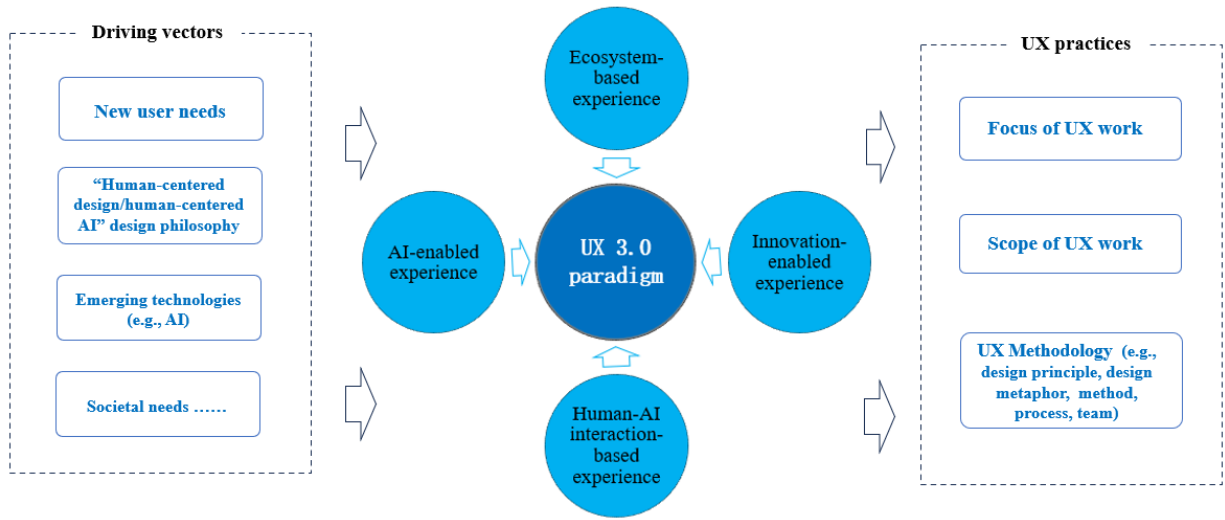

Figure 2  A conceptual framework of the UX.3.0 paradigm in the intelligence era (adapted from Xu, 2024)

As illustrated in Figure 2, the "UX 3.0" paradigm defines four emerging UX categories in support of designing and developing HCAI products in the intelligent era, including ecosystem-based experience, innovation-enabled experience, AI-enabled experience, and human-AI interaction-based experience. Overall, the "UX 3.0" paradigm framework represents the following strategies:

- **Advocating a systematic approach**: Consider UX in an all-around and systematic way. For example, the ecological perspective of experience, the perspective of experience-driven innovation;
- **Meeting emerging needs:** Consider the technologies and user needs in the intelligent era. For example, AI as a UX tool (i.e., AI-enabled experience), human-AI interaction-based experience in the intelligence era;
- **Guiding UX practice for HCAI products**: Define a series of UX methods to guide UX research and application to deliver human-centered AI products.

The " UX 3.0 " paradigm is significant to UX research and application. First, it helps UX professionals realize that the UX industry has entered a deep-water practice area in the intelligent era. UX research and application in the intelligent era have put forward new focus, scope, and methodology requirements. Existing UX practice cannot fully meet the needs and effectively address the new challenges posed by intelligent technology, we need to improve existing UX practice.

Secondly, it helps encourage UX professionals to carry out the necessary research and applications to improve existing UX methodology further. Entering the third stage of UX, improving UX paradigm, such as methodology, is one of the important topics in future UX research and application.

Finally, it helps UX professionals proactively strengthen collaboration with other related disciplines by leveraging their strengths and enhancing existing UX methodology. UX itself is an emerging field created through interdisciplinary collaboration. The emergence and methodology of the UX field more than 40 years ago were also driven by other disciplines. UX research and application in the intelligent era require more support from other disciplines.

## 3  Needs and implications for UX practice of the "UX 3.0" paradigm framework

## 3.1 Ecosystem-based experience

At present, users' experience is no longer limited to their experience based on the interactions with the UI of individual products at a time; it depends on the interactions with various ecosystems. Table 2 summarizes the four types of ecosystems and the emerging experiences from the interaction with these ecosystems. These emerging experiences drive new needs for the UX 3.0 paradigm and implications for UX practice. As shown in Table 2, these experiences are much broader, and we need new approaches to deliver seamless and end-to-end UX.

Table 2  Ecosystem-based experience

| Type | Emerging experiences of human-machine interaction | New needs for UX 3.0 | Implications for UX practice (examples) |
|---|---|---|---|
| Across the entire product life cycle | UX is beyond the interaction with a UI built during development and is expanded to the activities of pre-development (e.g., branding), development (e.g., UI, AI capabilities, business processes), and post-development (e.g., marketing, AI-enabled user support, product updates) | Optimize the interaction design of all the user touchpoints across the entire product life cycle, providing end-to-end UX | End-to-end UX (requirements, design, testing) methods, UX roadmap, etc. (Xu, 2014) |
| Across technological ecosystems | UX is beyond the interactions with individual products and is expanded to the interactions with and across technological ecosystems (e.g., operating systems, mobile/ ubiquitous computing), devices (e.g., desktop, phone, wearable), intelligent ecosystems (e.g., smart transportation, smart city), and online platforms (e.g., online shopping, social media) | Optimize the integration design within/across technological ecosystems, delivering seamless and consistent UX | End-to-end UX (requirements, design, testing) methods, UX roadmap, landing scenarios for emerging experiences, etc. |
| Across system architecture layers | UX is beyond the interaction with the product UI and is expanded to the interactions across the front-end (e.g., UI, AI capabilities), the middle-end (e.g., business logic and process), and the back-end (e.g., database, data quality, on-premises or cloud) | Optimize the integration design across the system architecture layers | UX architecture design, systematic integration design with interdisciplinary collaboration (Xu, Furie et al., 2019) |
| Within a broad sociotechnical environment | UX is beyond the micro context of interaction between users and products; it is expanded to the macro context of social, cultural, and organizational environments (e.g., smart city, smart transportation, smart factory). | Optimize the interaction between technical/AI and non-technical subsystems (e.g., social, cultural, organizational) in the sociotechnical environment | Intelligent sociotechnical systems methods (Xu, 2022b; Xu, Gao 2024) |

## 3.2 Innovation-enabled experience

Innovative design is already one of the driving forces for current social and economic development. Innovative design has become increasingly important with the increased homogeneity of products, agile development processes, and rising user expectations. At the same time, the value of "experience" has become increasingly prominent and has gradually become a key element in building differentiated and high-value competitive advantages.

Innovation is essentially a process of continuously adjusting user experience (user needs, usage scenarios, etc.) and technology to achieve the optimal human-machine relationship, making technology useful, easy to learn, and easy to use, thereby creating an innovative experience for humans (Evans et al., 2006; Xu, 2018). This "practical" innovation process is essentially UX-driven innovation. An increasing number of studies emphasize the contribution of UX to innovation (e.g., Luo, 2020). Previous innovation methods have their own pros and cons, but one thing in common is that they do not emphasize the role of UX in innovative design. People overemphasize the role of technology in driving innovative design and ignore the role of UX in the process, resulting in a relatively high failure rate (Yang et al., 2018; Debruyne, 2014; Li, 2017). Xu (2019) proposed a conceptual model of experience-driven innovation, emphasizing that innovation is essentially a UX-driven process. Table 3 summarizes the innovation-enabled experience and its implications for UX practice (Xu, 2018).

Table 3  Innovation-enabled experience

| Type | Innovation perspective | Implications for UX practice |
|---|---|---|
| User needs | Based on current user pain points | Use new technologies such as AI and big data to identify and address the user pain points among similar products, developing unique and differentiated experience solutions |

| | Based on potential user needs and usage scenarios | Use the new technologies to mine or predict potential (yet to be discovered or realized) valuable user needs supported by valid usage scenarios. |
|---|---|---|
| Technology-Enabled | Based on emerging interaction technology | Use the new technologies to develop innovative experiences (e.g., voice and gesture input) |
| | Based on existing technology | Use existing interaction technology (e.g., apply existing multiple single-channel interaction technology to address AR/VR motion problems with a multi-modal interaction approach) |
| | Based on real-time experience data | Use real-time data (e.g., user behavior, contextual data) to model digital personas and push for personalized functions and contents |
| Human-AI collaboration | Based on dynamic human-machine functional allocation | Based on the complementarity, dynamically adjust human-machine function and task allocation to achieve optimal system performance and experience |
| | Based on human-machine hybrid enhanced intelligence | Integrate human and machine intelligence to develop more powerful human-machine hybrid enhanced intelligence (e.g., the "human-in-the-loop" hybrid intelligence by leveraging the complementarity in intelligence) |
| | Based on human-machine teaming | Build effective human-AI partnerships for optimal system performance and experience based on the two-way and shared information, goals, tasks, execution, and decision-making |
| End-to-end experience | Based on sociotechnical systems approach | Deliver holistic end-to-end experience in the broad context of sociotechnical environment (e.g., innovative service design) |

All the methods listed in Table 3 reflect the interaction touch points between users and systems, products, or services at different levels. UX is generated from these interactive contacts. Therefore, these methods all reflect the design principle of experience-driven innovation. Also, these innovative design methods are actionable. In UX practice, UX professionals must work closely with other discipline teams to refine and improve these innovative methods.

**3.3 AI-enabled experience**

AI technology has revolutionized UX methods, and more and more AI technologies are being used in UX research and applications. AI-enabled experience can be achieved by supporting UX professionals' activities and providing real-time experience for users. From the perspective of UX professionals, AI technology has improved the work efficiency and design quality of UX professionals. From a user perspective, AI-enabled experience helps improve user experience; ChatGPT and Bard are good examples. Table 4 summarizes the "AI-enabled experience" methods for the "UX 3.0" paradigm framework (Xu, 2023a).

Table 4  AI-enabled experience

| Type | Implications for UX 3.0 | Design goals |
|---|---|---|
| AI-enabled user research | Intelligent collection of user insights (e.g., online data), intelligent analysis of user interview (e.g., voice content), intelligent qualitative analysis of user data, digital intelligent user portraits, digital intelligent customer journey map | • Enhance UX methods<br>• Improve UX activity efficiency<br>• Optimize experience design<br>• Enhance real-time user experience |
| AI-enabled UI design | AI generative UI design (Li et al., 2020), AI generative tools such as ChatGPT, Bard, Adobe Firefly, etc.), "AI as a new design material," collaborative UI design (De Peuter et al., 2021; Xu, 2023b) | |
| AI-enabled UX validation | Intelligent experience testing, user online data analysis, multi-modal (eye movement, expression, EEG, etc.) data collection/analysis (Chromik et al., 2020) | |
| AI-enabled real-time experience for users | Personalized function and content push based on real-time experience data, intelligent assistant systems (ChatGPT, etc.), intelligent recommendation systems, intelligent voice assistants | |

Overall, these methods are not yet fully mature. For example, AI design assistance tools for UI prototyping design value the novelty of appearance but lack attention to the UX design itself. It cannot provide the best UX for UX designers, such as building an architecture based on effective business processes and interactions. Moreover, the target users of some prototyping tools based on AI technology are mostly software developers, resulting in these tools not meeting the special needs of UX design professionals and the UX design process (Sun et al., 2020).

In the long run, AI technology can serve as an auxiliary tool for UX design to improve the work efficiency of UX designers. However, it may not completely replace creative UX design. UX professionals need to proactively work with AI professionals to optimize these methods based on the needs of UX activities.

**3.4 Human-AI interaction-based experience**

Traditional human-computer interaction (HCI) studies the interaction between humans and non-intelligent computing systems. HCI is currently transforming into the interaction with AI-based intelligent systems. This change brings a richer experience and puts forward new requirements for UX methods in the intelligent era (Ozmen Garibay et al., 2023).

AI technologies have unique autonomous characteristics; intelligent agents can have some human-like cognitive characteristics (e.g., self-learning, autonomous execution ) (Xu, 2020). On the one hand, these new features provide powerful technical support for developing natural and effective experiences; on the other hand, these new features also bring unique experiences, and previous UX methods based on non-intelligent computing systems cannot effectively solve these new problems. Table 5 summarizes the emerging features of interaction brought by intelligent technology, new needs for the UX 3.0 paradigm, and the implications for UX practice.

Table 5  Human-AI interaction-based experience

| Type | Emerging features of human-machine interaction | New needs for UX 3.0 | Implications for UX practice (examples) |
|---|---|---|---|
| **Optimized AI behavior** | • AI systems exhibit evolving machine behavior (Rahwan et al., 2019)<br>• Intelligent autonomous capabilities may handle scenarios unpredictable by design (Xu, 2020)<br>• AI machine behavior can lead to biased output (Abbas et al., 2022)<br>• Uncertainty and unpredictability of machine behavior impact UX | • Optimize machine behavior evolution and reduce algorithm bias<br>• Support experience of human-controllable AI<br>• Prevent users from overly trusting AI<br>• Improve users' ability to monitor and handle abnormal scenarios | • Apply user participation, iterative prototyping, and UX testing to optimize algorithm training and testing<br>• Collect user feedback to optimize AI system machine behavior evolution |
| **Explainable AI** | • AI "black box effect" can cause the unexplainable of AI system output (Yang et al., 2022 )<br>• AI explainability impacts on user trust and experience | • User-centered explainable AI design<br>• Verify the effectiveness of explainable AI designs | • Visualization design for transparent UI<br>• UI models of explainable AI<br>• Psychological explanation theory-based approaches |
| **Human-AI collaboration** | • Both humans and AI agents can collaborate as teammates<br>• Shared trust, mental models, situation awareness, decision-making, and controllability between humans and AI<br>• Best use of human and machine intelligence to maximize system performance and UX | • Human-AI collaboration brings in a new form of UX<br>• Collaboration-based cognitive user interface<br>• Dynamic task and functional allocation of humans and machines | • Modeling method to perceive, understand, and predict human-AI synergy state<br>• UX requirement definition, models, and methods for human-AI collaboration<br>• Collaborative UI models and design<br>• Methods for dynamic tasks/functional allocation<br>• Team performance and UX evaluation method for human-AI collaboration |
| **Intelligent UI** | • Both human and AI systems, as agents, can initiate interactions<br>• Interaction based on the sensing and reasoning of user intention and context (instead of "simple attributes" and "accurate input")<br>• Social and emotional interactions | • Human-machine interface that supports human-AI collaboration<br>• More natural and usable human-computer/AI interaction<br>• New interaction paradigms and UI design metaphors | • User state and intention recognition modeling method<br>• Intelligent human-AI interaction design standards (Amershi et al., 2019)<br>• Human-AI interface prototyping method<br>• Social and emotional interactive experience design methods |
| **Ethical AI** | • Emerging UX based on new user needs (e.g., privacy, fairness, ethics, skill growth, decision-making authority) beyond conventional needs for usability and functionality | • AI systems should reflect new user needs and corresponding experiences | • UX methods from a sociotechnical systems perspective (Xu, Gao, 2024)<br>• Ethical AI design based on interdisciplinary methods |

The changes like the human-AI teaming relationship in the intelligent era provide more possibilities for improving the experience of human-AI interaction and also put forward new requirements for experience design in the context of human-AI collaboration. The application of large AI language model systems such as ChatGPT-4 and Bard has begun to emphasize the idea of human-intelligence collaborative cooperation. Microsoft Office 365 Copilot system ). However, these systems are currently far from reaching optimal experience. UX professionals must fully understand this emerging human-machine relationship and explore how to use this human-intelligence collaborative relationship to enhance the user experience of using intelligent systems. UX professionals should actively cooperate with other disciplines, exploring experience design in the context of human-AI collaboration. These issues are also important for UX in the future intelligent era.

## 4  Future work for implementing the "UX 3.0" paradigm framework

First, establish new design thinking. UX professionals need to regard the intelligence era as an important opportunity to improve UX methodology. AI technology is more than just a tool. UX professionals need to transform their design thinking to provide effective experience solutions that meet user needs. For example, experience-driven innovative thinking for the "innovation-enabled experience" and the human-AI teaming metaphor for the "human-AI interaction-based experience."

Secondly, carry out research and application of the "UX 3.0" paradigm, such as design guidelines, process, method, design metaphor, etc. (Xu, Gao, and Dainoff, 2024). Research on the "UX 3.0" paradigm not only involves collaboration with other related disciplines but also requires further work on specific methods, UX evaluation and validation systems, UX process enhancement, and standardization.

Third, cultivate the talents of UX professionals. In the first and second stages, the UX field was basically a shallow practice area with low entry barriers. Entering the deep-water area of UX practice in the intelligent era (the third stage), the demand for talent in the UX industry is showing a rapid growth trend. At the same time, The practice of the "UX 3.0" paradigm also requires more specialized knowledge and skills. UX professionals need to improve their knowledge of intelligent technology, systematic methods, human factors science (e.g., engineering psychology, human factors engineering), etc. For students majoring in UX or students interested in UX, universities need to offer programs or courses such as "human-AI interaction" and "human-centered AI. In addition, through scientific research and postgraduate training, we will research the "UX 3.0" paradigm and methods to cultivate high-level UX professionals who meet the needs of the intelligence era.

Fourth, strengthen interdisciplinary collaboration. Perfecting the "UX 3.0" paradigm has exceeded the scope of the existing knowledge system in the UX field. It requires multidisciplinary collaboration, including engineering psychology, human factors engineering, HCI, AI, computer science, etc. At the same time, practicing the "UX 3.0" paradigm also requires the support of multidisciplinary collaboration (Xu, Dainoff, 2023).

Finally, optimize the process and organizational environment for designing and developing intelligent products. The effective application of the "UX 3.0" paradigm also depends on whether UX can be effectively integrated into product development processes, methods, and environments. At the project level, interdisciplinary collaboration should be maximized, and existing development processes need to be optimized based on the "user-centered " concept by establishing multidisciplinary project teams and adopting multidisciplinary methods. At the enterprise and organizational level, cultivate an organizational culture based on the "user-centered" design philosophy, formulate development standards based on emerging requirements of human-AI interaction.

## 5  Conclusion

Throughout its history of more than 30 years, UX has shown obvious staged development characteristics. Emerging technologies, user needs, and UX practices promote the evolvement of UX paradigms. Such an evolvement promotes the growth of the UX field. The UX field is moving towards the intelligence era, which has put forward a series of new requirements for UX practice.

To meet the new needs of UX practice in the intelligence era, this paper proposes a framework for the "UX 3.0" paradigm in the intelligence era. The framework defines emerging experiences and the corresponding UX methodology.  The "UX 3.0" paradigm helps UX professionals realize that the intelligence era has put forward new needs for UX practice. Existing UX practices cannot fully meet the needs of the intelligence era. We need to improve existing UX methodology, which is one of the important topics in future UX research and application. The UX field needs to evolve the UX paradigm in the intelligence era through multidisciplinary collaboration and methods.